\documentclass[12pt]{article}
\usepackage[koi8-ru]{inputenc}
\usepackage[russian]{babel}
\usepackage{epsf}
\usepackage{pazh}
\usepackage{amsmath}	
\usepackage{amssymb}	
\newcommand{\ergs}{\,erg\,s$^{-1}$}
\newcommand{\kms}{\,km\,s$^{-1}$}
\newcommand{\cms}{\,cm\,s$^{-1}$}
\newcommand{\gcmq}{\,g\,cm$^{-3}$}

\newcommand{\gcm}{\,g\,cm$^{-1}$}
\newcommand{\ha}{H$\alpha$}

\begin{document}
\title{\bf First day of type IIP supernova SN 2013fs: \ha\ 
 from preshock accelerated gas
}
\author{\bf \hspace{-1.3cm}\copyright\, 2020  \\
N. N. Chugai
}
\affil{
Institute of Astronomy of Russian Academy of Sciences, Moscow\\ 
}

\vspace{2mm}

\sloppypar 
\vspace{2mm}
\noindent


\noindent

\vfill
\noindent\rule{8cm}{1pt}\\
{$^*$ email $<$nchugai@inasan.ru$>$}

\clearpage

\centerline{\bf Abstract}
\vspace{0.5cm}

I explore the origin of an asymmetry of the \ha\ emission from a circumstellar (CS) 
  shell 
  around type IIP supernova SN~2013fs in the spectrum taken 10.3\,h after the shock breakout.
A spherical model of the \ha\ emission from the CS shell 
   that takes into account a preshock gas acceleration by 
  a supernova radiation permits us to successfully reproduce the \ha\ profile.
Principal factors responsible for the \ha\ asymmetry are the high 
   velocity of the accelerated CS preshock gas ($\sim 3000$\kms) and a low \ha\ 
   Sobolev optical depth in a combination with an occultation of the \ha\ emission 
   by the photosphere. 
  
\vspace{0.5cm}

\section{Introduction}

Type IIP supernova (SN~IIP) prior to the explosion has a red   
  supergiant (RSG) structure and loses its material via the slow wind (10-30\kms).
The supernova decceleration in the wind brings about radio and X-ray 
  emission 
  (Chevalier 1982) thus permitting us to infer the wind density parameter 
  $w = \dot{M}/u \sim 10^{14}-10^{15}$\gcm\ (Chevalier et al. 2006).
Optical emission lines from the wind of that density are too weak for 
their detection.
Particularly, the expected recombination \ha\ luminosity of the RSG wind is 
  $\approx 4.8\cdot10^{37}w_{15}^2/(v_9t_d)$\ergs, where $v_9$ is the maximal velocity 
  of the undisturbed supernova envelope in units of $10^9$\cms, $t_d$ is the 
   supernova age 
  in days, $w_{15}$ is in units of $10^{15}$\gcm. 
 Such a wind suggests the \ha\ luminosity $\sim 5\cdot10^{35-37}$\ergs\ and 
 the Thomson optical depth $\tau_{\sc T} \ \sim (0.03 - 0.3)/t_d$.
 
 Meanwhile, spectra of some SNe~IIP taken during first couple days after the explosion 
   show rather strong emission lines with narrow core and broad wings (Groh 2014, 
   Khazov et al. 2016, Yaron et al. 2017). 
 These lines indicate that the radiation of the circumstellar shell 
   experiences multiple Thomson scattering implying shell optical depth in the range of 
   2-3 (Groh 2014, Chugai 2001).   
Of special interest is well studied SN~2013fs (Yaron et al. 2017) with a set of spectra 
  taken starting with 6 h after the shock breakout.
The luminosity of \ha\ on 1.4 day from the CS shell is of $1.9\cdot10^{39}$\ergs, far above the expected luminosity from RSG wind.
Observational data suggest that preSN in that case has a dense CS shell with the radius of 
 $< 10^{15}$\,cm and mass of (several)$\times10^{-3}$\msun\ (Yaron et al. 2013).

The origin of a confined dense shell around SNe~IIP is not fully understood.
 The shell might originate due to the enhanced mass loss during several years prior 
   to the explosion.
 Alternatively, this shell could be a buffer zone between the RSG atmosphere 
   and the wind likewise the shell around Betelgeuse (Dessart et al. 2017).
 The latter extends up to $\sim 2\cdot10^{15}$\,cm and has aspherical clumpy structure 
  (Kervella et al. 2011).
Its dynamical equilibrium is maintained presumably by the RSG pulsations or/and 
  vigorous convection (Kervella et al. 2011).
This picture is supported by CO observations that see upward and downward motions 
 with velocities in the range 10--30\kms\ (Ohnaka et al. 2011).
 
The study of confined dense shell of SNe~IIP is crucial for the understanding its 
  origin, and particularly for the issue, to what extent 
  the dense shell in SN~IIP is of the same nature as the shell of Betelgeuse.
In this regard it has not escaped my attention the fact that the SN~2013fs spectrum 
  taken on Keck/LRIS at 10.3 hours after the shock breakout shows a pronounced asymmetry in \ha\ (Yaron et al. 2017) that is not explaned by the spectral model.
A hint at similar asymmetry is seen in previous four Keck spectra but in the last 
 spectrum of the set the asymmetry is most apparent, which is emphasised by 
 the Gaussian decomposition (Yaron et al. 2017, Supplement Fig. 2).
It is noteworthy that in the descibed model (Yaron et al. 2017) 
  \ha\ forms in the CS shell with the constant expansion velocity of 100\kms\  
  that is some approximation since a significant preshock acceleration 
  probably takes place at this early phase.

The present paper addresses the issue of the \ha\ asymmetry in the early 
   SN~2013fs spectrum hopefully to understand, whether this asymmetry could 
   originate in a spherically symmetric CS shell or may be it is related to 
   the asphericity of the CS shell.
 As will be shown, the \ha\ asymetry is not caused by the shell asymmetry, 
   instead it arises naturally in the spherical case.
  
The study is based on the SN~2013fs spectrum taken by Keck-I/LRIS 10.3 hours 
  after the shock breakout (Yaron et al. 2017). 
The spectrum is retrieved from the database  WISeREP (Yaron \& Gal-Yam 2012) 
  ({\em https://wiserep.weizmann.ac.il}).

\section{Modelling \ha}
\label{sec:mod}

\subsection{Model overview}

Radiative cooling of the shock wave following the shock breakout favours  
  the formation of a thin dense shell which serves as the photosphere for 
  several days in the case of SN~2013fs (Chugai 2020).
It should be emphasised that the photosphere is defined as the level at 
   which the effective optical depth is unity.
The CS shell even in the case of the Thomson optical depth of 
$\tau_{\sc T} \approx 2$ is effectively thin, i.e.,  
   $\tau_{\sc T}[k_a/(k_a+k_{\sc T})]^{1/2} \ll 1$, where
   $k_a$ and $k_{\sc T}$ are coefficients of the absorption and Thomson scattering respectively.
 The kinetic temperature in the CS shell is assumed to be equal to the photosphere 
 temperature 25000\,K with the photospheric radius of $R_1 = 10^{14}$\,cm at $t = 10.3$\,h  (Yaron et al. 2017).
The outer radius of the CS shell is adopted to be $R_2 = 5\cdot10^{14}$\,cm (Yaron et al. 2017). 
Following the model of early SN~1998S (Chugai 2001) we adopt that 
  the spherical photosphere of the radius $R_1$ with sharp boundary is imbedded into an ionized CS gas distributed in the range of $R_1 < r < R_2$.
The density distribution is assumed to be homogeneous in the range of 
 in accord with the previous model (Chugai 2020).  
   
 A powerful initial luminosity accelerates the CS gas, which 
   results in the formation before the outer shock of the velocity 
   distribution with the negative velocity gradient.
Basically, the kinematics of the accelerated CS gas should be computed by the 
  supernova explosion model (e.g., Dessart et al. 2017). 
However we use here a convenient description
\begin{equation}
 v = (v_1 - v_2)[(R_2 - r)/(R_2 - R_1)]^q + v_2\,,
 \end{equation}
   where $v_1$ is the gas velocity before the shock wave at $r = R_1$ 
   and $v_2$ is the velocity of the undisturbed wind at $R_2$ taken 
   to be 50\kms, close to the spectral resolution (60\kms).

 The free parameters are $\tau_{\sc T}$, $v_1$, $q$, and the parameter of the Sobolev 
   optical depth $\tau_0 = (\pi e^2/mc)f_{23}\lambda_{23}n_2(R_1/v_1)$, where 
   $f_{23}$ and $\lambda_{23}$ are the \ha\ oscilator strength and wavelength, 
   $n_2$ is the  population of the second level; the rest of values have usual sense.
 The Sobolev optical depth at the radius $r$ for a photon with the wave vector ${\bf k}$, 
  and a direction $\mu = ({\bf kv})/(kv)$ is 
  $\tau_{\sc S} = \tau_0(v_1/R_1)(r/v)/[1 - \mu(\gamma -1)]$, where 
  $\gamma = (r/v)(dv/dr)$.
 
 The radiation transfer in the CS shell with Thomson and resonant scatterings 
  is calculated emloying the Monte Carlo technique.
 The model includes also a diffuse reflection from the photosphere. 
 But this effect is small even for high albedo $\omega = 0.5$ since reflected 
   photons scatter in the far blue wing due to 
  the high photosphere velocity of $\sim 26000$\kms\ at this stage (Chugai 2020) 
  and do not affect the profile in the considered range of radial velocities, 
  so we adopt $\omega = 0$.

  \subsection{Results} 
  
The preliminary modelling suggests that the formerly adopted SN 2013fs redshift 
    $v_{rs} = 3554$\kms\ according to the redshift of the host galaxy NGC 7610 
    ($z = 0.011855$, NED) in fact should be corrected by the additional 
    redshift +140\kms.
The conclusion is supported by the spectrum taken on day 51 
  (P200/BBSP, Yaron et al. 2017).
The latter contains weak narrow emissions of \ha\ and [O\,III] 5006.843\,\AA,   
   both showing additional redshift +140\kms.
Below we use the corrected redshift  $v_{rs} = 3694$\kms.    

The modelling reveals that for any choice of parameters the agreement with the 
  observed spectrum requires optically thin \ha. 
For $\tau_0 > 0.1$ the observed line asymmetry cannot be reproduced.
The optimal model (Fig. 1) with parameters  $\tau_0 = 0$, $v_1 = 3000$\kms, 
 $\tau_{\sc T} = 2$, and $q = 1.7$ successfully describes the \ha\ profile 
 including the asymmetry that is most apparent in Fig. 1b.
Parameter uncertainties are $\pm 0.3$ for $\tau_{\sc T}$, $\pm 0.1$ for $q$, 
  and $\pm 500$\kms\ for $v_1$. 
  
The effect of the low \ha\ optical depth of $\tau_0 = 0.2$ for 
 the unchanged other parameters is shown in Fig. 2a.
In this case the model shows the prononced absorption and the asymmetry 
 of the opposite sense 
  with the red wing stronger than the blue one.
This is convincing demonstration that the \ha\ indeed should be optically thin.
The model with the preshock velocity $v_1 =1000$\kms\ (Fig. 2b) 
  demonstrates the significance of the preshock velocity.
In this case the asymmetry almost fully disappeares, which suggests that 
  the model with the higher preshock velocity is preferred.

The conclusion on the small \ha\ optical depth can be compared to  
  the computation of the hydrogen ionization and excitation of the second level.
The two-level ($n = 1$ and $n = 2$) plus continuum approximation is used 
  with all the relevant radiation and collisional processes taken into account.  
The stationary kinetic equations are solved for parameters $R_1$, $R_2$, and $T$
  adopted above for the optimal model and electron number density 
  suggested by the Thomson optical depth $\tau_{\sc T} = 2$.
The found solution shows high ionization $x = 0.999$ and low population of the 
  second level that implies the low Sobolev optical depth 
  $\tau_{\sc S} = 7\cdot10^{-4}$ at the level $r = R_1$ in accord with the \ha\ modelling.
A problem is however that the Sobolev optical depth increases outward and 
  at the radius $r = 3.5R_1$ becomes large $\tau_{\sc S} = 0.1$ in obvious 
  disagreement with the conclusion about transparent \ha.
The controversy can be resolved, if one admits that the shell is clumpy with the 
  volume filling factor $f < 0.01$.
It should be emphasised that the clumpiness does not affect the Thomson optical depth 
  provided the number of clouds along the shell radius exceeds unity.
      
 It is critical to check whether the radiation is able to accelerate CS gas 
   upto 3000\kms.
 A sensible estimate can be provided by the acceleration due to the Thomson scattering 
   and neglecting gas displacement during the acceleration time. 
 The equation of motion results in the velocity at the radius $r$ at the given age $t$ 
 \begin{equation}
   v = \frac{k_{\sc T}E_r}{4\pi r^2c} = 8.5\cdot10^7E_{r,48}r_{14}^{-2}~\mbox{\cms},  
 \label{eq:accel}  
 \end{equation}  
 where $k_{\sc T} = 0.34$\,cm$^2$\,g$^{-1}$ is the Thomson opacity,
  $E_{r,48}$ is the integrated luminosity during 10.3\,h in units of
   $10^{48}$\,erg, $r_{14}$ is the radius in units of $10^{14}$\,cm, and other 
   values have the usual sense.
 The SN~2013fs luminosity at $t = 3.6$ after the shock breakout is 
   $L \approx 3\cdot10^{44}$\ergs\ (Yaron et al. 2017) that implies the 
   radiation energy $E_r \approx 4\cdot10^{48}$\,erg.
With this value the equation (\ref{eq:accel}) gives the velocity estimate 
  of $\sim 3400$\kms\ at the radius of $10^{14}$\,cm and the moment of 10.3\,h.
The preshock velocity in our model 3000\kms\ thus turns out 
  to be consistent within uncertainties with the expected velocity of the gas 
  accelerated by the radiation.

\section{Discussion and Conclusions}

The aim of this paper has been to provide an explanation for the \ha\ asymmetry 
  in the early spectrum of SN 2013fs.
We find that the spherical model for the \ha\ emission in the dense CS shell 
  is able to account for the asymmetry provided special conditions are met. 
Specifically, apart from the Thomson scattering the model should include 
  the presence of the CS gas accelerated by the radiation. 
Remarkably, that the recovered preshock velocity of 3000\kms\ turns out to be 
  in good agreament with the value expected from the radiative acceleration. 
The additional requirement for the asymmetry is the very low Sobolev optical depth 
  in \ha, which takes place in the preshock layers of the CS gas. 
However the clumpy  structure with the low filling factor seems to be needed 
  in outer layers of the CS shell to avoid significant \ha\ absorption. 
In this regard it would be of great interest to check whether other 
 SNe~IIP with dense confined shells would reveal the  
 \ha\ absorption component in day-age spectra?

Dessart et al. (2017) computed spectra for the arbitrary SN~IIP model 
  at the early stages for 
  different cases of a dense CS envelope with the preshock kinematics  
  produced by the radiation hydrodynamics.
The reported model spectra show two cases with the required  \ha\ asymmetry: 
  the model r1w5r at 20\,h and the model r1w5h at 11.6\,h.
However it is not clear whether the employed computational procedure 
  would show the required \ha\ asymmetry in the model appropriate for 
   SN 2013fs at 10.3\,h. 
 
In the case of SN~1998S the model for the early \ha\ with the narrow core and 
 broad wings takes into account the radiative acceleration of the CS gas 
with the preshock velocity of 1000\kms\ (Chugai 2001).
However neither the observed line, nor the model profile show the line asymmetry. 
The reason is that due to the large Thomson optical depth 
 ($\tau_{\sc T} = 3.6$) and the low preshock velocity the Thomson scattering 
  washes out the asymmetry effect in this case.
 
 An interesting coincidence draws our attention. 
 For the density of the CS shell of $\rho \approx 1.5\cdot10^{-14}$\gcmq\ 
   implied by the optimal model and the velocity of the photosphere of 
   $v_s = 2.6\cdot10^4$\kms\ at 10.3\,h (Chugai 2020) the kinetic luminosity 
   of the shock wave is $L = 2\pi R_1^2\rho v_s^3 = 1.6\cdot10^{43}$\ergs\ that 
   coincides with the estimated observational bolometric luminosity at this moment 
  (Yaron et al. 2017).
  At first glance the outer shock determines the luminosity at this stage. 
 However, for the CS density in the optimal model the cooling time of the postshock 
   gas is several days, so a significant contribution of the forward shock to the 
   supernova luminosity at 10\,h is doubtful.
 
 The spherical symmetry of the confined dense shell of SN~2013fs raises 
   a question whether a close analogy between CS shell in SN 2013fs and the shell of Betelgeuse is actually takes place. 
 The point is that the shell around Betelgeuse  shows strong asphericity 
  and clumpiness (Kervella et al. 2011).
 If the latter asymmetry is caused by the asymmetry of the mass loss due to a
   large scale convection (Kervella et al. 2011) then one should admit that 
   the sphericity of shell around preSN~IIP indicates rather the spherical regime of 
  the mass loss by preSN probably due to radial pulsations.


\pagebreak   

\pagebreak   

\clearpage
\begin{figure}[h]
	\epsfxsize=19cm
	\hspace{-2cm}\epsffile{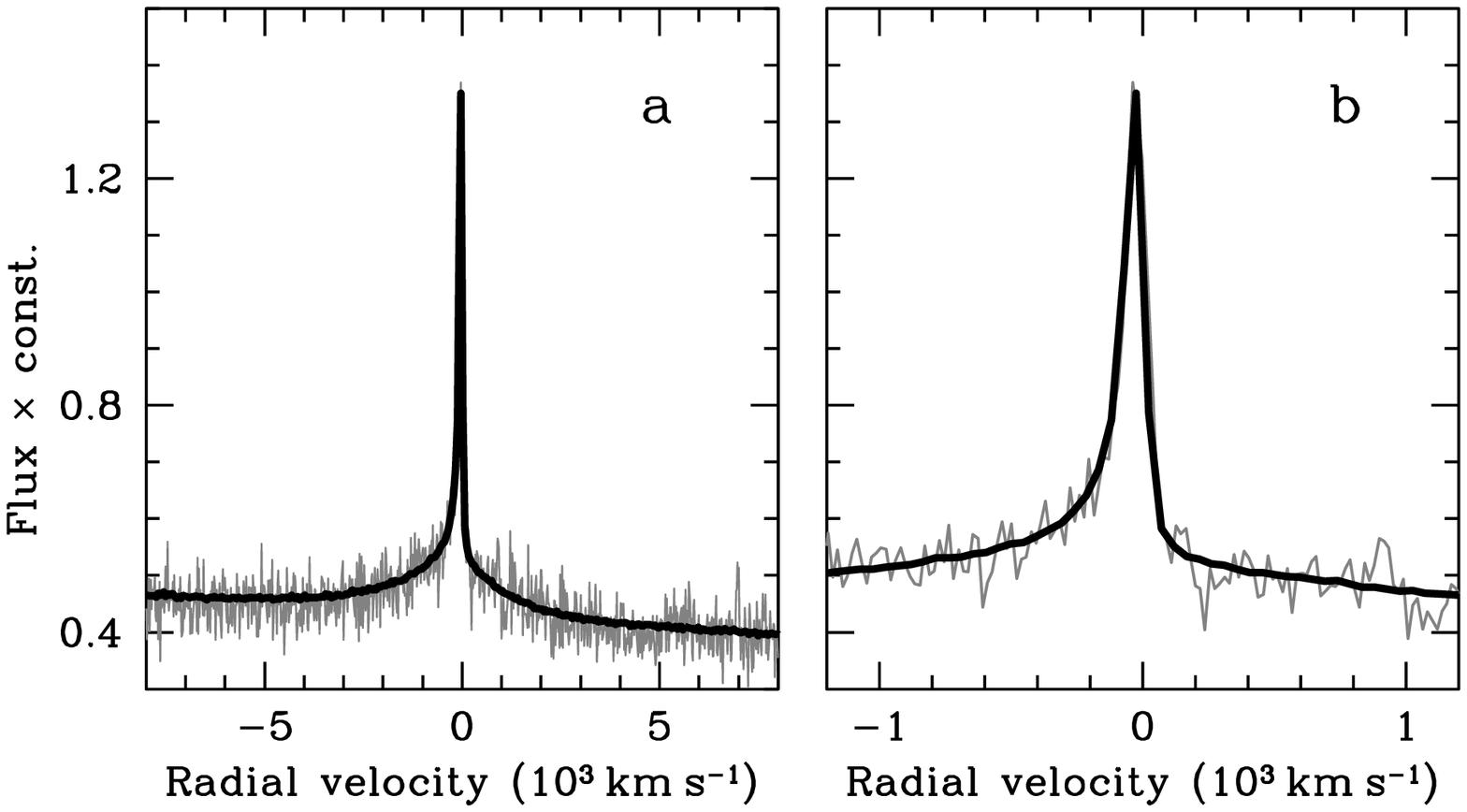}
	\caption{\rm  
	\ha\ in SN 2013fs spectrum at 10.3\,h after the shock breakout.
	The optimal model is overplotted on the observed spectrum ({\it grey}) in the 
	large range of radial velocities ({\bf a}) and close to the line center ({\bf b}). 
	In the latter figure the line asymmetry looks more obvious.
	}
\end{figure}
\clearpage
\begin{figure}[h]
	\epsfxsize=19cm
	\hspace{-2cm}\epsffile{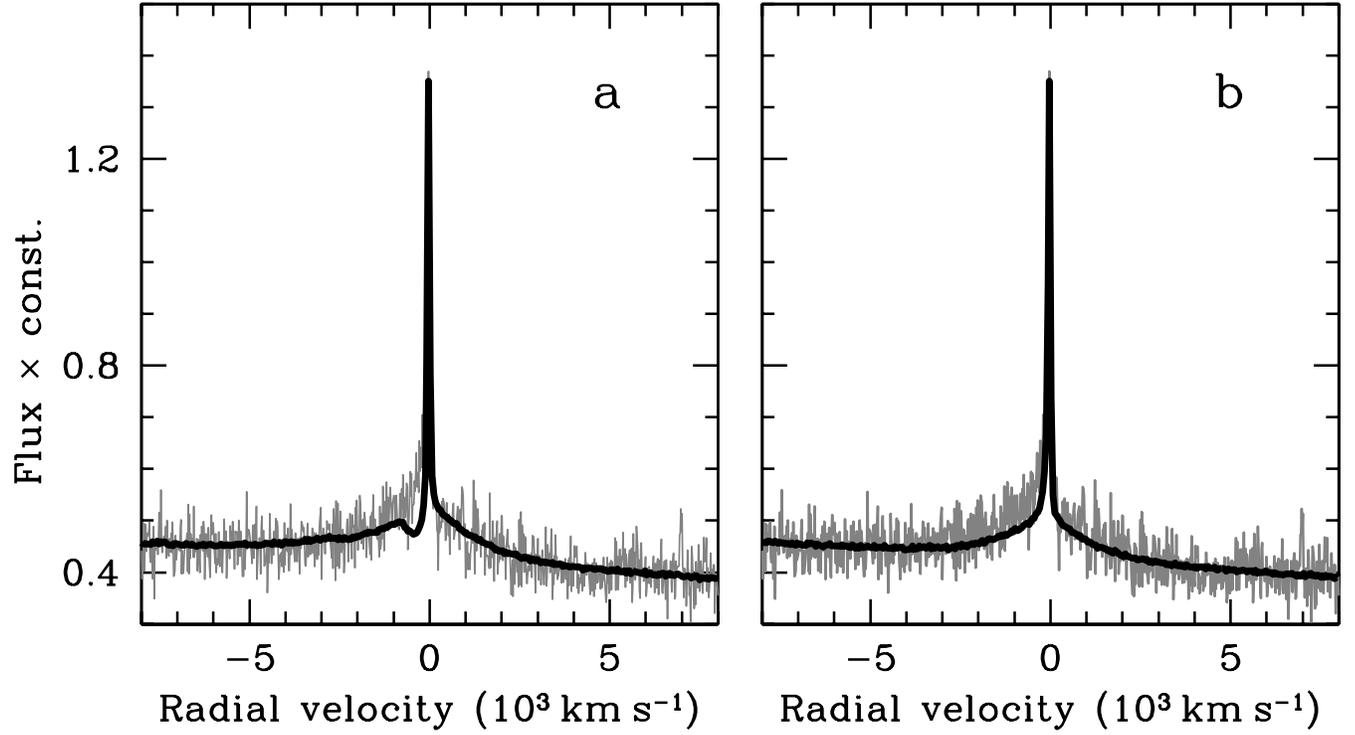}
	\caption{\rm  
	The same as Fig. 1, but Fig. 2a showing the optimal model in which the 
	optical depth parameter of \ha\ is $\tau_0 = 0.2$, and Fig. 2b 
	showing the model with the preshock velocity of 1000\kms. 
	Both models apparently are inconsistent with the observed spectrum.
		}
\end{figure}

\end{document}